\documentclass[]{article}
\usepackage{proceed2e}
\usepackage[protrusion=true,spacing=true]{microtype}
\usepackage{amsmath,amsfonts,amssymb,amsthm}
\usepackage[noend]{algorithmic}
\usepackage{algorithm}
\usepackage{graphicx,subfig}
\usepackage[round]{natbib}
\usepackage{times}
\usepackage{cite}
\usepackage{setspace}

\usepackage[absolute,overlay]{textpos}
\newtheoremstyle{mystyle}
  {5pt}		
  {0pt}		
  {}			
  {}			
  {\bfseries}	
  {.}			
  { }			
  {}			
\theoremstyle{mystyle}

\newcommand{\hy}{\pi_j^*}
\newcommand{\hysp}{\Pi_j^i}
\newcommand{\resinf}{s_{j-i}^t}
\newcommand{\delpi}{\delta^t(\hy)}
\newcommand{\ab}{\textbf{a}}
\newcommand{\at}{\ab_j^t}
\newcommand{\attil}{\tilde{\ab}_j^t}
\newcommand{\athat}{\hat{\ab}_j^t}
\newcommand{\atau}{\ab_j^\tau}
\newcommand{\atautil}{\tilde{\ab}_j^\tau}
\newcommand{\atauhat}{\hat{\ab}_j^\tau}
\newcommand{\ex}{\mathbb{E}}
\newcommand{\var}{\text{Var}}
\newcommand{\R}{\mathbb{R}}
\newcommand{\tighteq}{\hspace{-4pt}=\hspace{-4pt}}
\newcommand{\vtighteq}{\hspace{-8pt}=\hspace{-8pt}}
\newcommand{\lheight}{0.09\textheight}
\newcommand{\lraise}{12pt}
\newcommand{\legpad}{\hspace{-460pt}}

\def\lf{\left\lfloor}   
\def\rf{\right\rfloor}

\title{Are You Doing What I Think You Are Doing? \\[2pt] Criticising Uncertain Agent Models}

\author{
	\textbf{Stefano V. Albrecht}\hspace{1pt} \\
	School of Informatics \\
	University of Edinburgh \\
	Edinburgh EH8 9AB, UK \\
	\texttt{s.v.albrecht@sms.ed.ac.uk}
	\And
	\hspace{1pt}\textbf{Subramanian Ramamoorthy} \\
	School of Informatics \\
	University of Edinburgh \\
	Edinburgh EH8 9AB, UK \\
	\texttt{s.ramamoorthy@ed.ac.uk}
}

\begin{document}

	\maketitle

	\begin{abstract}
The key for effective interaction in many multiagent applications is to reason explicitly about the behaviour of other agents, in the form of a \emph{hypothesised} behaviour. While there exist several methods for the construction of a behavioural hypothesis, there is currently no universal theory which would allow an agent to contemplate the correctness of a hypothesis. In this work, we present a novel algorithm which decides this question in the form of a frequentist hypothesis test. The algorithm allows for multiple metrics in the construction of the test statistic and learns its distribution during the interaction process, with asymptotic correctness guarantees. We present results from a comprehensive set of experiments, demonstrating that the algorithm achieves high accuracy and scalability at low computational costs.
	\end{abstract}

	\section{INTRODUCTION} \label{sec:intro}

A common difficulty in many multiagent systems is the fact that the behaviour of other agents may be initially unknown. Important examples include adaptive user interfaces, robotic elderly assistance, and electronic markets. Often, the key for effective interaction in such systems is to reason explicitly about the behaviour of other agents, typically in the form of a \emph{hypothesised} behaviour which makes predictions about future actions based on a given interaction history.

A number of methods have been studied for the construction of behavioural hypotheses. One method is to use opponent modelling techniques to learn a behaviour from the interaction history. Two well-known examples include fictitious play \citep{b1951} and case-based reasoning \citep{gs2001}, as well as their many variants. Another method is to maintain a set of possible action policies, called types, over which a posterior belief is computed based on the interaction history \citep{ar2014,gd2005}. The hypothesis is then obtained by using the posterior to mix the types. Related methods have been studied in the plan recognition literature \citep{c2001,cg1993}.

The learned behaviours (or models) of these methods can be viewed as hypotheses because they are eventually either true or false (subject to the various assumptions they are based on), and because they are \emph{testable}. Thus, the following is a natural question: given an interaction history $H$ and a hypothesis $\pi^*$ for the behaviour of an agent, does the agent indeed behave according to $\pi^*$? There are several ways in which an answer to this question could be utilised. For instance, if we persistently reject the hypothesis $\pi^*$, we may construct an alternative hypothesis or resort to some default plan of action (such as a ``maximin'' strategy).

Unfortunately, the above methods for hypothesis construction do not provide an answer to this question. Some opponent modelling methods use goodness-of-fit measures (e.g. those that rely on maximum likelihood estimation), but these measures describe how well the model fits the data (i.e. interaction history) and not necessarily if the model is correct. Similarly, the posterior belief in the type-based approach quantifies the relative likelihood of types (relative to a set of alternative types) but not the \emph{correctness} of types.

To illustrate the source of difficulty, consider the below excerpt of an interaction process between two agents which can choose from three actions. The columns show, respectively, the current time $t$ of the interaction, the actions chosen by the agents at time $t$, and agent 1's hypothesised probabilities with which agent $2$ would choose its actions at time $t$, based on the prior interaction history.

\begin{center}
	\begin{tabular}{ccc}
		$t$ & $(a_1^t,a_2^t)$ & $\pi_2^*$ \\
		\hline
		1 & $(1, 2)$ & $\langle .3, .1, .6 \rangle$ \\
		2 & $(3, 1)$ & $\langle .2, .3, .5 \rangle$ \\
		3 & $(2, 3)$ & $\langle .7, .1, .2 \rangle$ \\
		4 & $(2, 3)$ & $\langle .0, .4, .6 \rangle$ \\
		5 & $(1, 2)$ & $\langle .4, .2, .4 \rangle$
	\end{tabular}
\end{center}

Assuming that the process continues in this fashion, and without any restrictions on the behaviour of agent 2, how should agent 1 decide whether or not to reject its hypothesis about the behaviour of agent 2?

A natural way to address this question is to compute some kind of \emph{score} from the information given in the above table, and to compare this score with some manually chosen rejecting threshold. A prominent example of such a score is the empirical frequency distribution \citep{cs2007,fy2003}. While the simplicity of this method is appealing, there are two significant problems: (1) it is far from trivial to devise a scoring scheme that reliably quantifies ``correctness'' of hypotheses (for instance, an empirical frequency distribution taken over all past actions would be insufficient in the above example since the hypothesised action distributions are changing), and (2) it is unclear how one should choose the threshold parameter for any given scoring scheme.

In this work, we present an efficient algorithm which decides this question in the form of a frequentist hypothesis test. The algorithm addresses (1) by allowing for multiple scoring criteria in the construction of the test statistic, with the intent of obtaining an overall more reliable scoring scheme. The distribution of the test statistic is then learned during the interaction process, and we show that the learning is asymptotically correct. Finally, analogous to standard frequentist testing, the hypothesis is rejected at a given point in time if the resulting $p$-value is below some ``significance level''. This eliminates (2) by providing a uniform semantic for rejection that is invariant to the employed scoring scheme. We present a comprehensive set of experiments, demonstrating that our algorithm achieves high accuracy and scalability at low computational costs.

Of course, there is a long-standing debate on the role of statistical hypothesis tests and quantities such as $p$-values \citep[e.g.][]{gs2013,bs1987,c1977}. The usual consensus is that $p$-values should be combined with other forms of evidence to reach a final conclusion \citep{f1935}, and this is the view we adopt as well. In this sense, our method may be used as part of a larger machinery to decide the truth of a hypothesis.

	\section{RELATED WORK} \label{sec:relwork}

In addition to the related works mentioned in the previous section, there are a number of other related research areas:

There exists a large body of literature on what is often referred to as \emph{model criticism} \citep[e.g.][]{bb2000,m1994,r1984,b1980}. Model criticism attempts to answer the following question: given a data set $D$ and model $M$, could $D$ have been generated by $M$? This is analogous to our question, in which $D$ is a sequence of observed actions of some agent and $M$ is a hypothesised behaviour for that agent. However, in contrast to our work, model criticism usually assumes that the data are independent and identically distributed, which is not the case in the interactive settings we consider.

A related problem, sometimes referred to as \emph{\text{identity testing}}, is to test if a given sequence of data was generated by some given stochastic process \citep{rr2008,bs1977}. Instead of independent and identical distributions, this line of work assumes other properties such as stationarity and ergodicity. Unfortunately, these assumptions are also unlikely in interaction processes, and the proposed solutions are very costly.

Model criticism and identity testing are not to be confused with \emph{model selection}, in which two or more alternative models are under consideration \citep[e.g.][]{vo2012}. Similarly, we do not consider alternative hypotheses. However, our method can be applied individually to multiple hypotheses, or the hypotheses may be fused into a single hypothesis using a posterior belief \citep{ar2014,gd2005}.

Another related problem is that of \emph{model checking}, which attempts to verify that a given system (or model) satisfies certain formal properties \citep{c1999}. Recently, \citet{ar2014} applied the concept of probabilistic bisimulation \citep{ls1991} to the question of ``incorrect'' hypotheses and showed that a certain form of optimality is preserved if a bisimulation relation exists. However, their work is not concerned with establishing whether or not a given behavioural hypothesis is correct, and their analysis is performed \emph{before} any interaction.

Our method can be viewed as \emph{passive} in the sense that it does not actively probe different aspects of the hypothesis, and we show in Section~\ref{sec:exp} that this can be a drawback. This is in contrast to methods such as \citep{cm1999}, which promote active exploration. However, this exploration comes at high computational costs and limits the structure of hypotheses, such as deterministic finite state machines. On the other hand, our method has low computational costs and leaves the structure of the hypothesis open.

	\section{PRELIMINARIES} \label{sec:prel}

We consider a sequential interaction process with $m$ agents. The process begins at time $t = 0$. At each time $t$, each agent $i \in \left\{ 1,...,m \right\}$ receives a signal $s_i^t$ and chooses an action $a_i^t$ from a finite set of actions $A_i$. (Agents choose actions simultaneously.) The process continues in this fashion indefinitely or until some termination criterion is satisfied.

The signal $s_i^t$ specifies information that agent $i$ receives at time $t$ and may in general be the result of a random variable over past actions and signals. For example, $s_i^t$ may be a discrete system state and its dynamics may be described by some stochastic transition function. Note that we allow for asymmetric information (i.e. $s_i^t \neq s_j^t$). For example, $s_i^t$ may include a private payoff for agent $i$. In this work, we leave the precise structure and dynamics of $s_i^t$ open.

We assume that each agent $i$ can choose actions $a_i^t$ based on the entire interaction history $H_i^t = (s_i^0,a^0,s_i^1,a^1,...,s_i^t)$, where $a^\tau = (a_1^\tau,...,a_m^\tau)$ is the tuple of actions taken by the agents at time $\tau$. Formally, each agent $i$ has a \emph{behaviour} $\pi_i$ which assigns a probability distribution over actions $A_i$ given a history $H_i^t$, denoted $\pi_i(H_i^t)$. We use $\Pi_i$ to denote the infinite and uncountable space of all such behaviours. Note that a behaviour may implement any kind of logic, and it is useful to think of it as a black-box programme.

Given two agents $i$ and $j$, we use $\hysp$ to denote $i$'s \emph{hypothesis space} for $j$'s behaviours. The difference between $\hysp$ and $\Pi_j$ is that $\hy \in \hysp$ are defined over $H_i^t$ while $\pi_j \in \Pi_j$ are defined over $H_j^t$. Since we allow for asymmetric information, any information that is contained in $s_j^t$ but not in $s_i^t$, denoted $\resinf$, becomes part of the hypothesis space $\hysp$. For example, if $\resinf$ contains a private payoff for $j$, $i$ can hypothesise a payoff as part of its hypothesis for $j$'s behaviour.

Defining a behavioural hypothesis $\hy \in \hysp$ as a function $\hy(H_i^t)$ has two implicit assumptions: firstly, it assumes knowledge of $A_j$, and secondly, it assumes that the information in $\resinf$ is a (deterministic) function of $H_i^t$. If, on the other hand, we allowed $\resinf$ to be stochastic (i.e. a random variable over the interaction history), we would in addition have to hypothesise the random outcome of $\resinf$. In other words, $\hy(H_i^t)$ would itself be a random variable, which is outside the scope of this work.

	\section{A METHOD FOR BEHAVIOURAL HYPOTHESIS TESTING} \label{sec:main}

Let $i$ denote our agent and let $j$ denote another agent. Moreover, let $\hy \in \hysp$ denote our hypothesis for $j$'s behaviour and let $\pi_j \in \Pi_j$ denote $j$'s true behaviour. The central question we ask is if $\hy = \pi_j$?

Unfortunately, since we do not know $\pi_j$, we cannot directly answer this question. However, at each time $t$, we know $j$'s past actions $\at = (a_j^0,...,a_j^{t-1})$ which were generated by $\pi_j$. If we use $\hy$ to generate a vector $\athat = (\hat{a}_j^0,...,\hat{a}_j^{t-1})$, where $\hat{a}_j^\tau$ is sampled using $\hy(H_i^\tau)$, we can formulate the related two-sample problem of whether $\at$ and $\athat$ were generated from the same behaviour, namely $\hy$.

In this section, we propose a general and efficient algorithm to decide this problem. At its core, the algorithm computes a frequentist $p$-value
\begin{equation} \label{eq:p}
	p = P \left( |T(\attil,\athat)| \geq |T(\at,\athat)| \right)
\end{equation}
where $\attil \sim \delpi = (\hy(H_i^0),...,\hy(H_i^{t-1}))$. The value of $p$ corresponds to the probability with which we expect to observe a test statistic at least as extreme as $T(\at,\athat)$, under the null-hypothesis $\hy = \pi_j$. Thus, we reject $\hy$ if $p$ is below some ``significance level'' $\alpha$.

In the following subsections, we describe the test statistic $T$ and its asymptotic properties, and how our algorithm learns the distribution of $T(\attil,\athat)$. A summary of the algorithm is given in Algorithm~\ref{alg}.

\begin{algorithm}[t]
	\small
	\begin{spacing}{1.4}
		\begin{algorithmic}[1]
			\STATE \textbf{Input:} history $H_i^t$ (including observed action $a_j^{t-1}$)
			\STATE \textbf{Output:} $p$-value (reject $\hy$ if $p$ below some threshold $\alpha$)
			\STATE \textbf{Parameters:} hypothesis $\hy$; score functions $z_1,...,z_K$; $N\hspace{-2pt}>\hspace{-2pt}0$
			\STATE \textit{// Expand action vectors}
			\STATE Set $\ab_j^t \gets \langle \ab_j^{t-1} , a_j^{t-1} \rangle$
			\STATE Sample $\hat{a}_j^{t-1} \sim \hy(H_i^{t-1})$; set $\hat{\ab}_j^t \gets \langle \hat{\ab}_j^{t-1} , \hat{a}_j^{t-1} \rangle$
			\FOR{$n = 1,...,N$}
				\STATE Sample $\tilde{a}_j^{t-1} \sim \hy(H_i^{t-1})$; set $\tilde{\ab}_j^{t,n} \gets \langle \tilde{\ab}_j^{t-1,n} , \tilde{a}_j^{t-1} \rangle$
			\ENDFOR
			\STATE \textit{// Fit skew-normal distribution $f$}
			\IF{update parameters?}
				\STATE Compute $D \gets \left\{ T(\tilde{\ab}_j^{t,n},\athat) \ | \ n=1,...,N \right\}$
				\STATE Fit $\xi,\omega,\beta$ to $D$, \ e.g. using \eqref{eq:neglog}
				\STATE Find mode $\mu$ from $\xi,\omega,\beta$
			\ENDIF
			\STATE \textit{// Compute $p$-value}
			\STATE Compute $q \gets T(\at,\athat)$ using \eqref{eq:test}/\eqref{eq:tmulti}
			\RETURN $p \gets f(q \ | \ \xi,\omega,\beta) \ / \ f(\mu \ | \ \xi,\omega,\beta)$
			\vspace{-5pt}
		\end{algorithmic}
	\end{spacing}
	\caption{\small }
	\label{alg}
\end{algorithm}

		\subsection{TEST STATISTIC} \label{sec:test}

We follow the general approach outlined in Section~\ref{sec:intro} by which we compute a \emph{score} from a vector of actions and their hypothesised distributions. Formally, we define a \emph{score function} as $z : (A_j)^t \times \Delta(A_j)^t \rightarrow \R$, where $\Delta(A_j)$ is the set of all probability distributions over $A_j$. Thus, $z(\at,\delpi)$ is the score for observed actions $\at$ and hypothesised distributions $\delpi$, and we sometimes abbreviate this to $z(\at,\hy)$. We use $Z$ to denote the space of all score functions.

Given a score function $z$, we define the test statistic $T$ as \\[-15pt]
\begin{eqnarray} \label{eq:test}
	T(\attil,\athat) & \tighteq & \frac{1}{t} \sum_{\tau=1}^{t} T_\tau(\atautil,\atauhat) \\[3pt]
	T_\tau(\atautil,\atauhat) & \tighteq & z(\atautil,\hy) - z(\atauhat,\hy)
\end{eqnarray}

\vspace{-10pt}

where $\atautil$ and $\atauhat$ are the $\tau$-prefixes of $\attil$ and $\athat$, respectively.

In this work, we assume that $z$ is provided by the user. While formally unnecessary (in the sense that our analysis does not require it), we find it a useful design guideline to interpret a score as a kind of likelihood, such that higher scores suggest higher likelihood of $\hy$ being correct. Under this interpretation, a minimum requirement for $z$ should be that it is \emph{consistent}, such that, for any $t > 0$ and $\hy \in \hysp$,
\begin{equation} \label{eq:cons}
	\hy \in \Pi^z = \arg\max_{\pi_j' \in \hysp} \ex_{\ab_j' \sim \delpi} \left[ z(\ab_j',\pi_j') \right]
\end{equation}
where $\ex_\eta$ denotes the expectation under $\eta$. This ensures that if the null-hypothesis $\hy = \pi_j$ is true, then the score $z(\at,\hy)$ is maximised on expectation.

Ideally, we would like a score function $z$ which is \emph{perfect} in that it is consistent and $|\Pi^z| = 1$. This means that $\hy$ can maximise $z(\at,\hy)$ (where $\at \sim \delta^t(\pi_j)$) \emph{only} if $\hy = \pi_j$. Unfortunately, it is unclear if such a score function exists for the general case and how it should look. Even if we restrict the behaviours agents may exhibit, it can still be difficult to find a perfect score function. On the other hand, it is a relatively simple task to specify a small set of score functions $z_1,...,z_K$ which are consistent but imperfect. (Examples are given in Section~\ref{sec:exp}.) Given that these score functions are consistent, we know that the cardinality $|\cap_k \Pi^{z_k}|$ can only monotonically decrease. Therefore, it seems a reasonable approach to combine multiple imperfect score functions in an attempt to approximate a perfect score function.

Of course, we could simply define $z$ as a linear (or otherwise) combination of $z_1,...,z_K$. However, this approach is at risk of losing information from the individual scores, e.g. due to commutativity and other properties of the combination. Thus, we instead propose to compare the scores individually. Given score functions $z_1,...,z_K \in Z$ which are all bounded by the same interval $[a,b] \subset \R$, we redefine $T_\tau$ to
\begin{equation} \label{eq:tmulti}
	T_\tau(\atautil,\atauhat) = \sum_{k=1}^{K} w_k \left( z_k(\atautil,\hy) - z_k(\atauhat,\hy) \right)
\end{equation}
where $w_k \in \R$ is a weight for score function $z_k$. In this work, we set $w_k = \frac{1}{K}$. (We also experiment with alternative weighting schemes in Section~\ref{sec:exp}.) However, we believe that $w_k$ may serve as an interface for useful modifications of our algorithm. For example, \citet{yue2010} compute weights to increase the power of their specific hypothesis tests.

		\subsection{ASYMPTOTIC PROPERTIES} \label{sec:asym}

The vectors $\at$ and $\athat$ are constructed iteratively. That is, at time $t$, we observe agent $j$'s past action $a_j^{t-1}$, which was generated from $\pi_j(H_j^{t-1})$, and set $\ab_j^t = \langle \ab_j^{t-1} , a_j^{t-1} \rangle$. At the same time, we sample an action $\hat{a}_j^{t-1}$ using $\hy(H_i^{t-1})$ and set $\hat{\ab}_j^t = \langle \hat{\ab}_j^{t-1} , \hat{a}_j^{t-1} \rangle$. Assuming the null-hypothesis $\hy = \pi_j$, will $T(\at,\athat)$ converge in the process?

Unfortunately, $T$ might not converge. This may seem surprising at first glance given that $a_j^{t-1}, \hat{a}_j^{t-1}$ have the same distribution $\pi_j(H_j^{t-1}) = \hy(H_i^{t-1})$, since $\ex_{x,y \sim \psi} \left[ x-y \right] = 0$ for any distribution $\psi$. However, there is a subtle but important difference: while $a_j^{t-1},\hat{a}_j^{t-1}$ have the same distribution, $z_k(\at,\hy)$ and $z_k(\athat,\hy)$ may have arbitrarily different distributions. This is because these scores may depend on the entire prefix vectors $\ab_j^{t-1}$ and $\hat{\ab}_j^{t-1}$, respectively, which means that their distributions may be different if $\ab_j^{t-1} \neq \hat{\ab}_j^{t-1}$. Fortunately, our algorithm does not require $T$ to converge because it learns the distribution of $T$ during the interaction process, as we will discuss in Section~\ref{sec:learn}.

Interestingly, while $T$ may not converge, it can be shown that the fluctuation of $T$ is eventually normally distributed, for any set of score functions $z_1,...,z_K$ with bound $[a,b]$. Formally, let $\ex[T_\tau(\atau,\atauhat)]$ and $\var[T_\tau(\atau,\atauhat)]$ denote the finite expectation and variance of $T_\tau(\atau,\atauhat)$, where it is irrelevant if $\atau,\atauhat$ are sampled directly from $\delta^\tau(\hy)$ or generated iteratively as prescribed above. Furthermore, let $\sigma_t^2 \hspace{-1pt}=\hspace{-1pt} \sum_{\tau=1}^t \hspace{-1pt} \var[T_\tau(\atau,\atauhat)]$ denote the cumulative variance. Then, the standardised stochastic sum
\begin{equation}
	\frac{1}{\sigma_t} \sum_{\tau=1}^t T_\tau(\atau,\atauhat) - \ex[T_\tau(\atau,\atauhat)]
\end{equation}
will converge in distribution to the standard normal distribution as $t \rightarrow \infty$. Thus, $T$ is normally distributed as well.

To see this, first recall that the standard central limit theorem requires the random variables $T_\tau$ to be independent and identically distributed. In our case, $T_\tau$ are independent in that the random outcome of $T_\tau$ has no effect on the outcome of $T_{\tau'}$. However, $T_\tau$ and $T_{\tau'}$ depend on different action sequences, and may therefore have different distributions. Hence, we have to show an additional property, commonly known as \emph{Lyapunov's condition} \citep[e.g.][]{f2010}, which states that there exists a positive integer $d$ such that
\begin{equation}\label{eq:lyapunov}
	\lim_{t \rightarrow \infty} \frac{\hat{\sigma}_t^{2+d}}{\sigma_t^{2+d}} = 0 \text{, \ with}
\end{equation}
\vspace{-5pt}
\begin{equation}\label{eq:lyap-top}
	\hat{\sigma}_t^{2+d} = \sum_{\tau=1}^t \ex \left[ \left| T_\tau(\atau,\atauhat) - \ex[T_\tau(\atau,\atauhat)] \right|^{2+d} \right].
\end{equation}

Since $z_k$ are bounded, we know that $T_\tau$ are bounded. Hence, the summands in \eqref{eq:lyap-top} are uniformly bounded, say by $U$ for brevity. Setting $d = 1$, we obtain
\begin{equation}
	\lim_{t \rightarrow \infty} \ \frac{\hat{\sigma}_t^3}{\sigma_t^3} \ \leq \ \frac{U \hat{\sigma}_t^2}{\sigma_t^3} \ = \ \frac{U}{\sigma_t}
\end{equation}
The last part goes to zero if $\sigma_t \rightarrow \infty$, and hence Lyapunov's condition holds. If, on the other hand, $\sigma_t$ converges, then this means that the variance of $T_\tau$ is zero from some point onward (or that it has an appropriate convergence to zero). In this case, $\hy$ will prescribe deterministic action choices for agent $j$, and a statistical analysis is no longer necessary.

		\subsection{LEARNING THE TEST DISTRIBUTION} \label{sec:learn}

Given that $T$ is eventually normal, it may seem reasonable to compute \eqref{eq:p} using a normal distribution whose parameters are fitted during the interaction. However, this fails to recognise that the distribution of $T$ is shaped \emph{gradually} over an extended time period, and that the fluctuation around $T$ can be heavily skewed in either direction until convergence to a normal distribution emerges. Thus, a normal distribution may be a poor fit during this shaping period.

What is needed is a distribution which can represent any normal distribution, and which is flexible enough to faithfully represent the gradual shaping. One distribution which has these properties is the \emph{skew-normal distribution} \citep{a1985,ol1976}. Given the PDF $\phi$ and CDF $\Phi$ of the standard normal distribution, the skew-normal PDF is defined as
\begin{equation}
	f(x \ | \ \xi,\omega,\beta) = \frac{2}{\omega} \, \phi \left( \frac{x - \xi}{\omega} \right) \Phi \left( \beta \left( \frac{x - \xi}{\omega} \right) \right)
\end{equation}
where $\xi \in \R$ is the location parameter, $\omega \in \R^+$ is the scale parameter, and $\beta \in \R$ is the shape parameter. Note that this reduces to the normal PDF for $\beta = 0$, in which case $\xi$ and $\omega$ correspond to the mean and standard deviation, respectively. Hence, the normal distribution is a sub-class of the skew-normal distribution.

Our algorithm learns the shifting parameters of $f$ during the interaction process, using a simple but effective sampling procedure. Essentially, we use $\hy$ to iteratively generate $N$ additional action vectors $\tilde{\ab}_j^{t,1},...,\tilde{\ab}_j^{t,N}$ in the exact same way as $\athat$. The vectors $\tilde{\ab}_j^{t,n}$ are then mapped into data points
\begin{equation}
	D = \left\{ T(\tilde{\ab}_j^{t,n},\athat) \ | \ n=1,...,N \right\}
\end{equation}
which are used to estimate the parameters $\xi,\omega,\beta$ by minimising the negative log-likelihood
\begin{equation}\label{eq:neglog}
	N \log(\omega) - \sum_{x \in D} \log \phi \left( \frac{x - \xi}{\omega} \right) + \log \Phi \left( \beta \left( \frac{x - \xi}{\omega} \right) \right)
\end{equation}
whilst ensuring that $\omega$ is positive. An alternative is the method-of-moments estimator, which can also be used to obtain initial values for \eqref{eq:neglog}. Note that it is usually unnecessary to estimate the parameters at every point in time. Rather, it seems reasonable to update the parameters less frequently as the amount of evidence (i.e. observed actions) grows.

Given the asymmetry of the skew-normal distribution, the semantics of ``as extreme as'' in \eqref{eq:p} may no longer be obvious (e.g. is this with respect to the mean or mode?). In addition, the usual tail-area calculation of the $p$-value requires the CDF, but there is no closed form for the skew-normal CDF and approximating it is rather cumbersome. To circumvent these issues, we approximate the $p$-value as
\begin{equation}
	p \ \approx \ \frac{f(T(\at,\athat) \ | \ \xi,\omega,\beta)}{f(\mu \ | \ \xi,\omega,\beta)}
\end{equation}
where $\mu$ is the mode of the fitted skew-normal distribution. This avoids the asymmetry issue and is easier to compute.

\begin{figure*}[t]
	\includegraphics[width=1.00\textwidth]{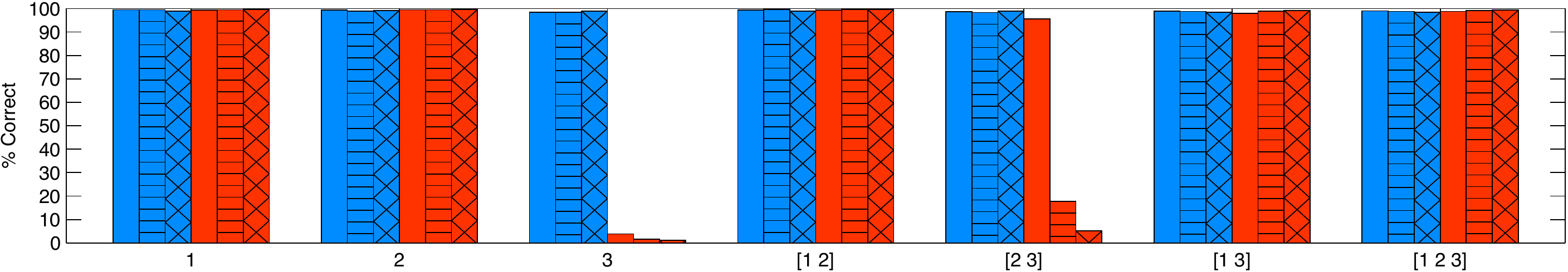}\legpad\raisebox{\lraise}{\includegraphics[height=\lheight]{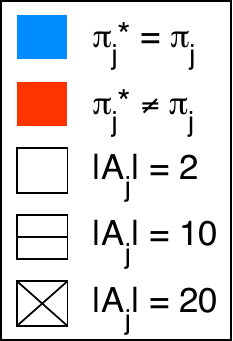}}
	\caption{Average accuracy with random behaviours, for $N=50$ and $|A_j|=2,10,20$. Results averaged over 500 processes with 10000 time steps, for $\hy = \pi_j$ and $\hy \neq \pi_j$ each. X-axis shows score functions $z_k$ used in test statistic.}
	\label{fig:ra-a21020-h50}
\end{figure*}

\begin{figure*}[t]
	\hspace{10pt}
	\subfloat[$|A_j| = 2$]{\raisebox{2pt}{\includegraphics[height=0.147\textheight]{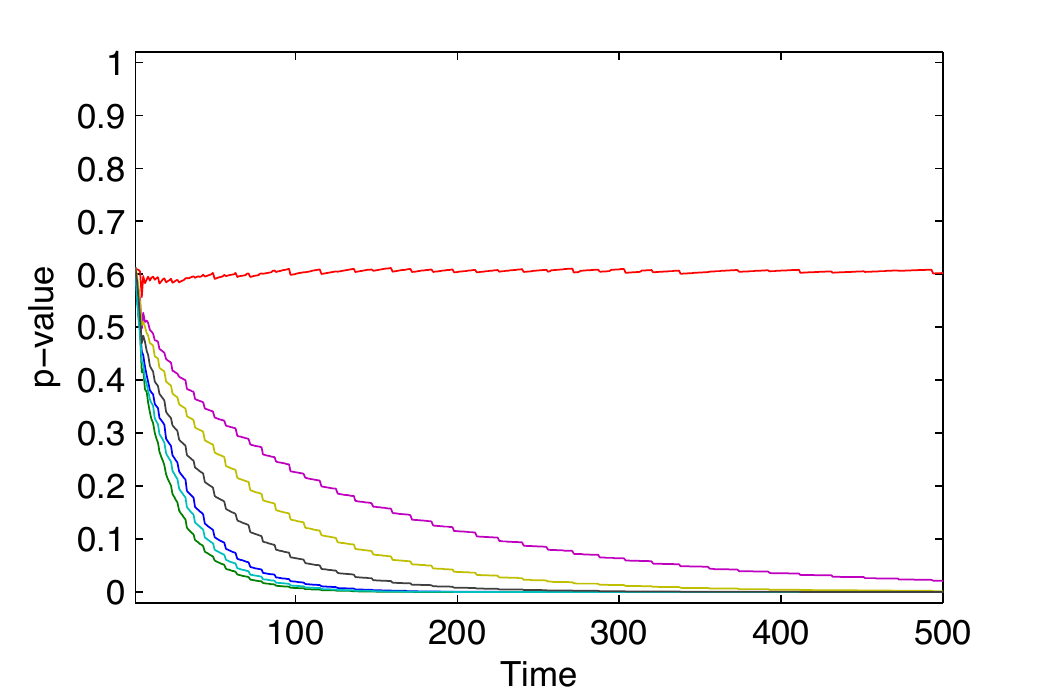}}\hspace{-50pt}\raisebox{40pt}{\includegraphics[height=0.08\textheight]{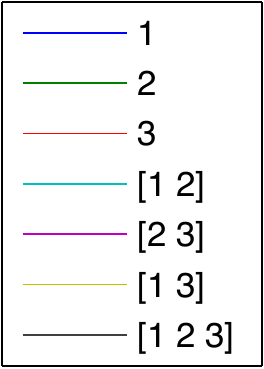}}}
	\hspace{30pt}
	\subfloat[$|A_j| = 10$]{\includegraphics[height=0.15\textheight]{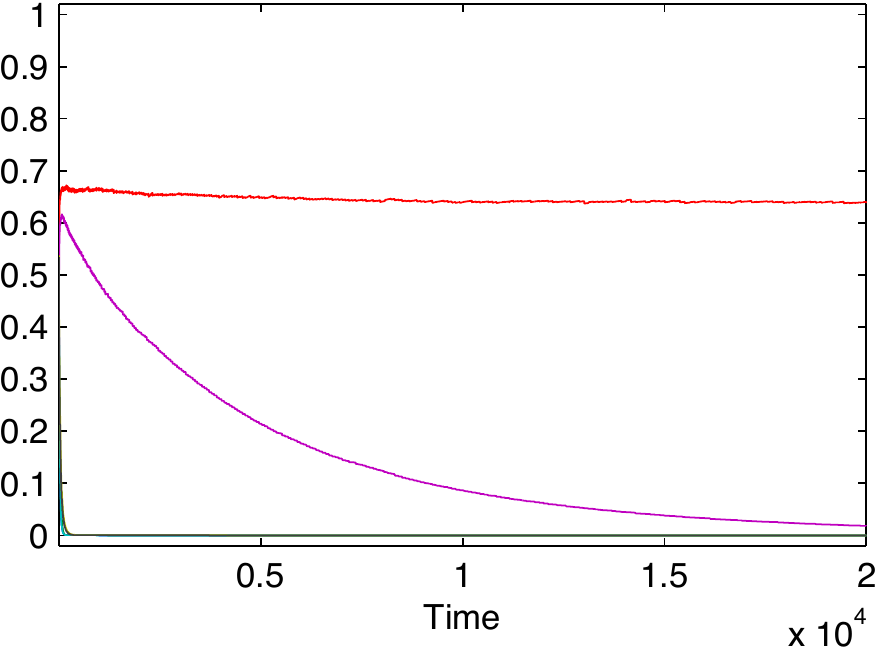}}
	\hspace{20pt}
	\subfloat[$|A_j| = 20$]{\includegraphics[height=0.15\textheight]{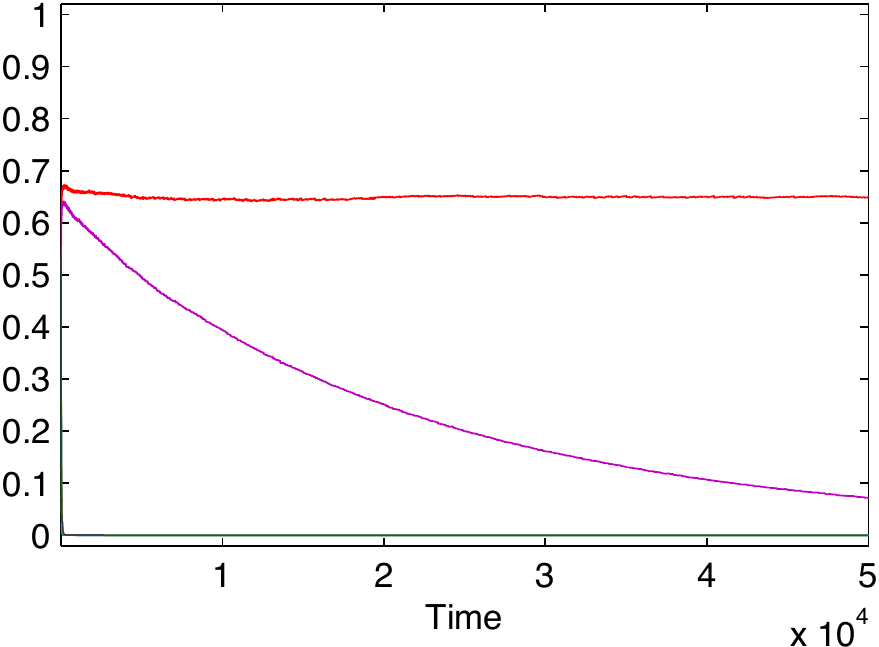}}
	\caption{Average $p$-values with random behaviours, for $N=50$ and $\hy \neq \pi_j$ (i.e. hypothesis wrong). Results averaged over 500 processes. Legend shows score functions $z_k$ used in test statistic.}
	\label{fig:ra-pval}
\end{figure*}

	\section{EXPERIMENTS} \label{sec:exp}

We conducted a comprehensive set of experiments to investigate the accuracy (correct and incorrect rejection), scalability (with number of actions), and sampling complexity of our algorithm. The following three score functions and their combinations were used:
\begin{eqnarray*}
	z_1(\at,\hy) & \vtighteq & \frac{1}{t} \sum_{\tau=0}^{t-1} \frac{\hy(H_i^\tau)[a_j^\tau]}{\max_{a_j \in A_j} \hy(H_i^\tau)[a_j]} \\[2pt]
	z_2(\at,\hy) & \vtighteq & \frac{1}{t} \sum_{\tau=0}^{t-1} \hspace{-1pt} 1 \hspace{-1pt}-\hspace{-1pt} \ex^{a_j \sim}_{\hy(H_i^\tau)} \hspace{-2pt} \left| \hy(H_i^\tau)[a_j^\tau] \hspace{-1pt}-\hspace{-1pt} \hy(H_i^\tau)[a_j] \hspace{-1pt} \right| \\[2pt]
	z_3(\at,\hy) & \vtighteq & \hspace{-4pt} \sum_{a_j \in A_j} \hspace{-5pt} \min \hspace{-2pt} \left[ \frac{1}{t} \sum_{\tau=0}^{t-1} [a_j^\tau \hspace{-1pt}=\hspace{-1pt} a_j]_1 , \frac{1}{t} \sum_{\tau=0}^{t-1} \hy(H_i^\tau)[a_j] \right]
\end{eqnarray*}
where $[b]_1 = 1$ if $b$ is true and 0 otherwise. Note that $z_1,z_3$ are generally consistent (cf. Section~\ref{sec:test}), while $z_2$ is consistent for $|A_j| = 2$ but not necessarily for $|A_j| > 2$. Furthermore, $z_1,z_2,z_3$ are all imperfect. The score function $z_3$ is based on the empirical frequency distribution (cf. Section~\ref{sec:intro}).

The parameters of the test distribution (cf. Section~\ref{sec:learn}) were estimated less frequently as $t$ increased. The first estimation was performed at time $t = 1$ (i.e. after observing one action). After estimating the parameters at time $t$, we waited $\lf \sqrt{t} \rf - 1$ time steps until the parameters were re-fitted. Throughout our experiments, we used a significance level of $\alpha = 0.01$ (i.e. reject $\hy$ if the $p$-value is below $0.01$).

		\subsection{RANDOM BEHAVIOURS}

In the first set of experiments, the behaviour spaces $\Pi_i,\Pi_j$ and hypothesis space $\hysp$ were restricted to ``random'' behaviours. Each random behaviour is defined by a sequence of random probability distributions over $A_j$. The distributions are created by drawing uniform random numbers from $(0,1)$ for each action $a_j \in A_j$, and subsequent normalisation so that the values sum up to 1.

Random behaviours are a good baseline for our experiments because they are usually hard to distinguish. This is due to the fact that the entire set $A_j$ is always in the support of the behaviours, and since they do not react to any past actions. These properties mean that there is little structure in the interaction that can be used to distinguish behaviours.

We simulated 1000 interaction processes, each lasting 10000 time steps. In each process, we randomly sampled behaviours $\pi_i \in \Pi_i,\ \pi_j \in \Pi_j$ to control agents $i$ and $j$, respectively. In half of these processes, we used a correct hypothesis $\hy = \pi_j$. In the other half, we sampled a random hypothesis $\hy \in \hysp$ with $\hy \neq \pi_j$. We repeated each set of simulations for $|A_j| = 2,10,20$ (with $|A_i| = |A_j|$) and $N = 10,50,100$ (cf. Section~\ref{sec:learn}).

			\subsubsection{Accuracy \& Scalability}

Figure~\ref{fig:ra-a21020-h50} shows the average accuracy of our algorithm (for $N = 50$), by which we mean the average percentage of time steps in which the algorithm made correct decisions (i.e. no reject if $\hy = \pi_j$; reject if $\hy \neq \pi_j$). The x-axis shows the combination of score functions used to compute the test statistic (e.g. [1 2] means that we combined $z_1,z_2$).

The results show that our algorithm achieved excellent accuracy, often bordering the 100\% mark. They also show that the algorithm scaled well with the number of actions, with no degradation in accuracy. However, there were two exceptions to these observations: Firstly, using $z_3$ resulted in very poor accuracy for $\hy \neq \pi_j$. Secondly, the combination of $z_2,z_3$ scaled badly for $\hy \neq \pi_j$.

The reason for both of these exceptions is that $z_3$ is not a good scoring scheme for random behaviours. The function $z_3$ quantifies a similarity between the empirical frequency distribution and the averaged hypothesised distributions. For random behaviours (as defined in this work), both of these distributions will converge to the uniform distribution. Thus, under $z_3$, any two random behaviours will eventually be the same, which explains the low accuracy for $\hy \neq \pi_j$.

As can be seen in Figure~\ref{fig:ra-a21020-h50}, the inadequacy of $z_3$ is solved when adding any of the other score functions $z_1,z_2$. These functions add discriminative information to the test statistic, which technically means that the cardinality $|\Pi^z|$ in \eqref{eq:cons} is reduced. However, in the case of $[z_2,z_3]$, the converge is substantially slower for higher $|A_j|$, meaning that more evidence is needed until $\hy$ can be rejected. Figure~\ref{fig:ra-pval} shows how a higher number of actions affects the average convergence rate of $p$-values computed with $z_2,z_3$.

In addition to the score functions $z_k$, a central aspect for the convergence of $p$-values are the corresponding weights $w_k$ (cf. \eqref{eq:tmulti}). As mentioned in Section~\ref{sec:test}, we use uniform weights $w_k = \frac{1}{K}$. However, to show that the weighting is no trivial matter, we repeated our experiments with four alternative weighting schemes: Let $z^\tau_k= z_k(\atautil,\hy) - z_k(\atauhat,\hy)$ denote the summands in \eqref{eq:tmulti}. The weighting schemes \texttt{truemax}\,/\,\texttt{truemin} assign $w_k = 1$ for the first $k$ that maximises\,/\,minimises $|z^\tau_k|$, and 0 otherwise. Similarly, the weighting schemes \texttt{max}\,/\,\texttt{min} assign $w_k = 1$ for the first $k$ that maximises\,/\,minimises $z^\tau_k$, and 0 otherwise.

\begin{figure*}[t]
	\includegraphics[width=1.00\textwidth]{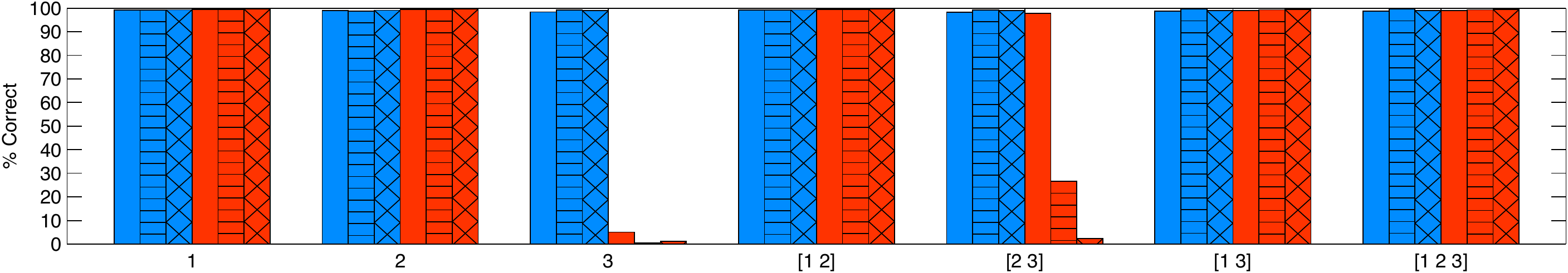}\legpad\raisebox{\lraise}{\includegraphics[height=\lheight]{legend_A.pdf}}
	\caption{Average accuracy with random behaviours, for $N=50$ and $|A_j|=2,10,20$. X-axis shows score functions $z_k$ used in test statistic. Weights $w_k$ computed using \texttt{truemax} weighting.}
	\label{fig:ra-a21020-h50-truemax}
\end{figure*}

\begin{figure*}[t]
	\includegraphics[width=1.00\textwidth]{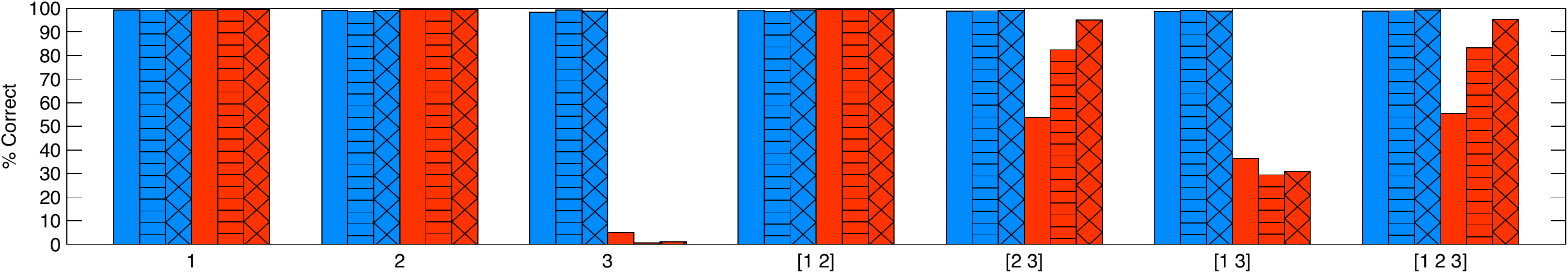}\legpad\raisebox{\lraise}{\includegraphics[height=\lheight]{legend_A.pdf}}
	\caption{Average accuracy with random behaviours, for $N=50$ and $|A_j|=2,10,20$. X-axis shows score functions $z_k$ used in test statistic. Weights $w_k$ computed using \texttt{truemin} weighting.}
	\label{fig:ra-a21020-h50-truemin}
\end{figure*}

Figures~\ref{fig:ra-a21020-h50-truemax} and \ref{fig:ra-a21020-h50-truemin} show the results for \texttt{truemax} and \texttt{truemin}. As can be seen in the figures, \texttt{truemax} is very similar to uniform weights while \texttt{truemin} improves the convergence for $[z_2,z_3]$ but compromises elsewhere. The results for \texttt{max} and \texttt{min} are very similar to those of \texttt{truemin} and \texttt{truemax}, respectively, hence we omit them.


Finally, we recomputed all accuracies using a more lenient significance level of $\alpha = 0.05$. As could be expected, this marginally decreased and increased (i.e. by a few percentage points) the accuracy for $\hy = \pi_j$ and $\hy \neq \pi_j$, respectively. Overall, however, the results were very similar to those obtained with $\alpha = 0.01$. 

			\subsubsection{Sampling Complexity}

Recall that $N$ specifies the number of sampled action vectors $\tilde{\ab}_j^{t,n}$ used to learn the distribution of the test statistic (cf. Section~\ref{sec:learn}). In the previous section, we reported results for $N=50$. In this section, we investigate differences in accuracy for $N=10,50,100$.

\begin{figure}[t]
	\centering
	\includegraphics[height=0.18\textheight]{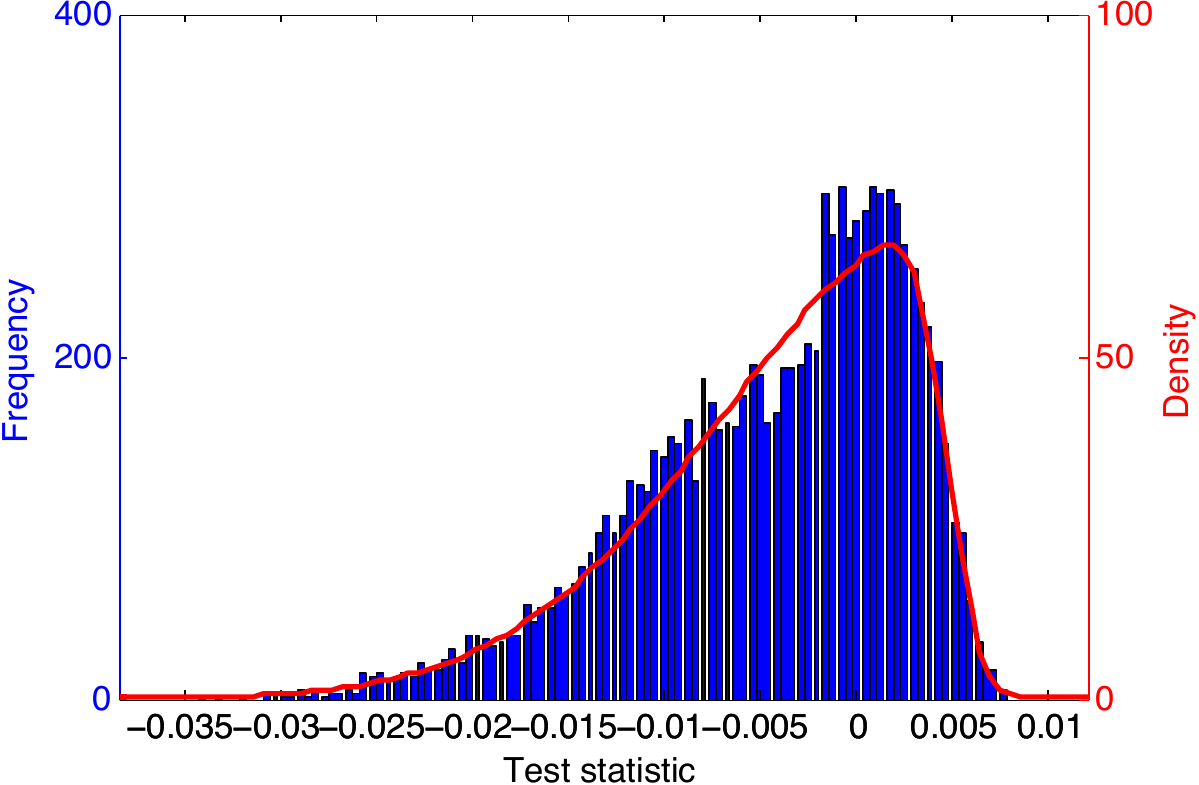}
	\caption{True test distribution for $z_2$ (histogram) and learned skew-normal distribution (red curve) after 10 time steps, with $|A_j|=10$ and $N=100$.}
	\label{fig:ra-hist-skew}
\end{figure}

\begin{figure*}[t]
	\includegraphics[width=1.00\textwidth]{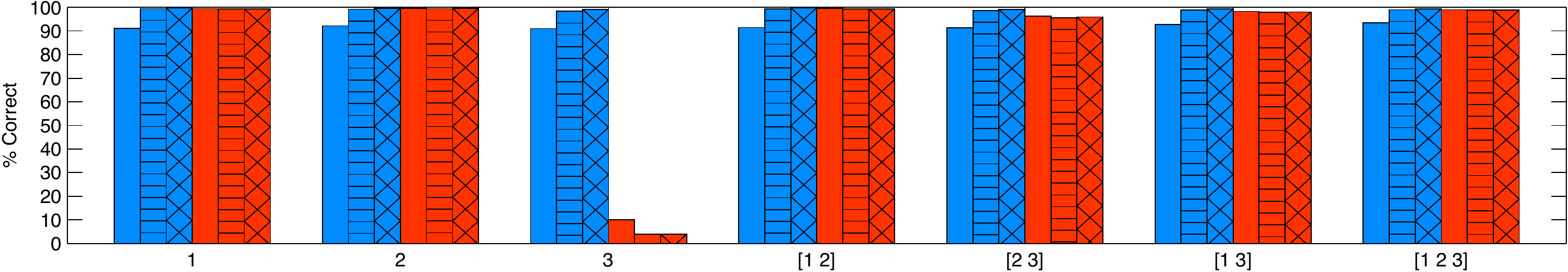}\legpad\raisebox{\lraise}{\includegraphics[height=\lheight]{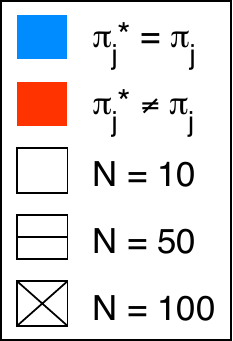}}
	\caption{Average accuracy with random behaviours, for $|A_j|=2$ and $N=10,50,100$. Results averaged over 500 processes with 10000 time steps, for $\hy = \pi_j$ and $\hy \neq \pi_j$ each. X-axis shows score functions $z_k$ used in test statistic.}
	\label{fig:ra-a2-h1050100}
\end{figure*}

\begin{figure*}[t]
	\includegraphics[width=1.00\textwidth]{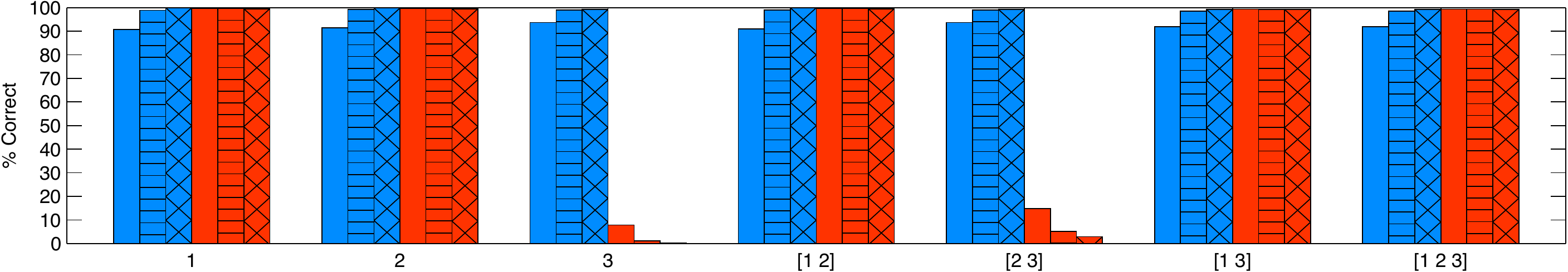}\legpad\raisebox{\lraise}{\includegraphics[height=\lheight]{legend_N.pdf}}
	\caption{Average accuracy with random behaviours, for $|A_j|=20$ and $N=10,50,100$. Results averaged over 500 processes with 10000 time steps, for $\hy = \pi_j$ and $\hy \neq \pi_j$ each. X-axis shows score functions $z_k$ used in test statistic.}
	\label{fig:ra-a20-h1050100}
\end{figure*}

Figures~\ref{fig:ra-a2-h1050100} and \ref{fig:ra-a20-h1050100} show the differences for $|A_j| = 2,20$, respectively. (The figure for $|A_j| = 10$ was virtually the same as the one for $|A_j| = 20$, except with minor improvements in accuracy for the $[z_2,z_3]$ cluster. Hence, we omit it here.) As can be seen, there were improvements of up to 10\% from $N=10$ to $N=50$, and no (or very marginal) improvements from $N=50$ to $N=100$. This was observed for all $|A_j|=2,10,20$, and all constellations of score functions. The fact that $N=50$ was sufficient even for $|A_j|=20$ is remarkable, since, under random behaviours, there are $20^t$ possible action vectors to sample at any time $t$.

\begin{figure}[t]
	\centering
	\includegraphics[height=0.135\textheight]{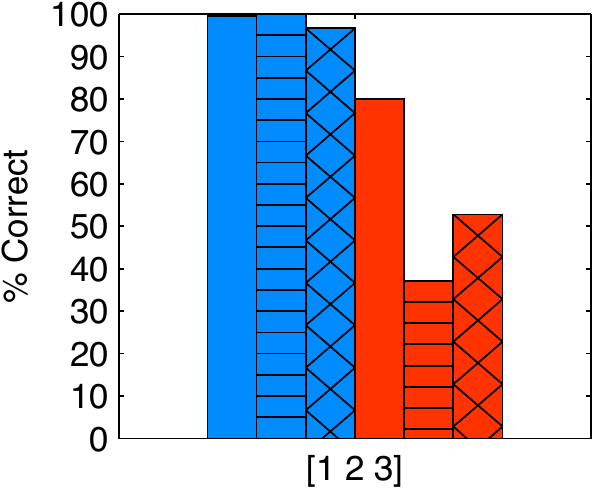}\hspace{15pt}\raisebox{18pt}{\includegraphics[height=\lheight]{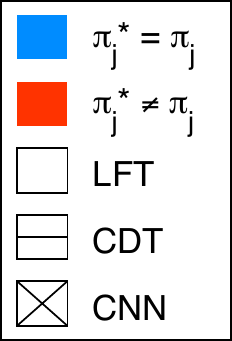}}
	\caption{Average accuracy for behaviour classes LFT, CDT, CNN ($N = 50$). $\Pi_i$ and $\Pi_j$ restricted to same class.}
	\label{fig:lcc-a0}
\end{figure}
\begin{figure}[t]
	\centering
	\includegraphics[height=0.135\textheight]{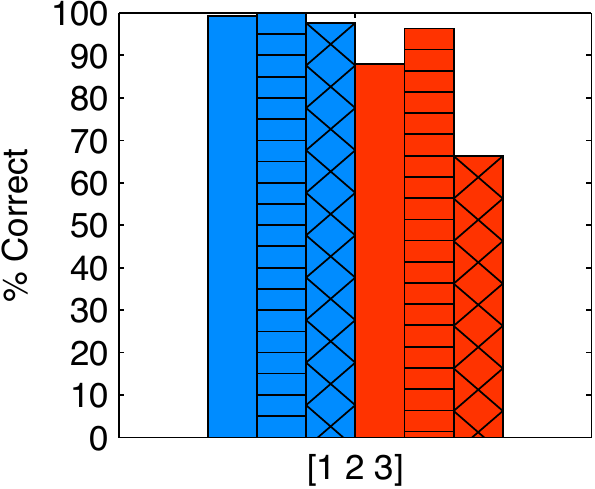}\hspace{15pt}\raisebox{18pt}{\includegraphics[height=\lheight]{legend_lcc.pdf}}
	\caption{Average accuracy for behaviour classes LFT, CDT, CNN ($N = 50$). $\Pi_i$ set to random behaviours.}
	\label{fig:lcc-a1}
\end{figure}

\begin{figure*}[t]
	\vspace{-8pt}
	\centering
	\subfloat[$N = 10$]{\includegraphics[height=0.153\textheight]{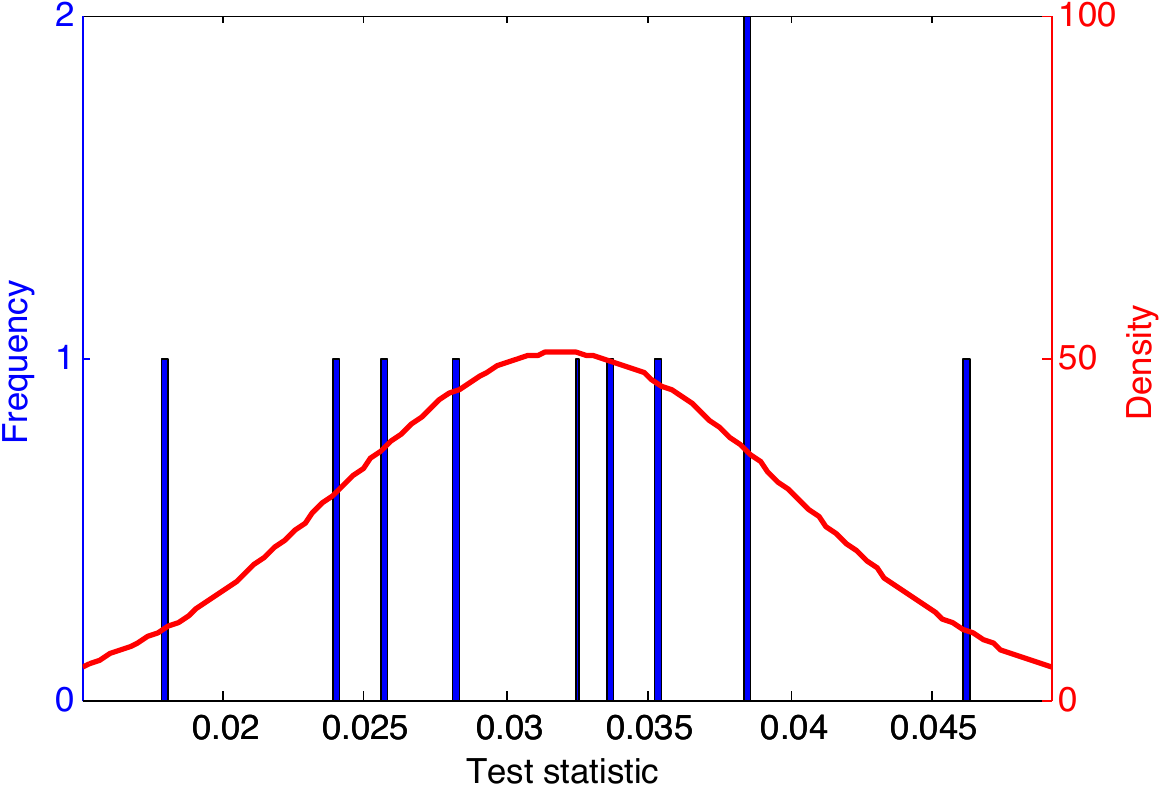}}
	\hspace{10pt}
	\subfloat[$N = 50$]{\includegraphics[height=0.153\textheight]{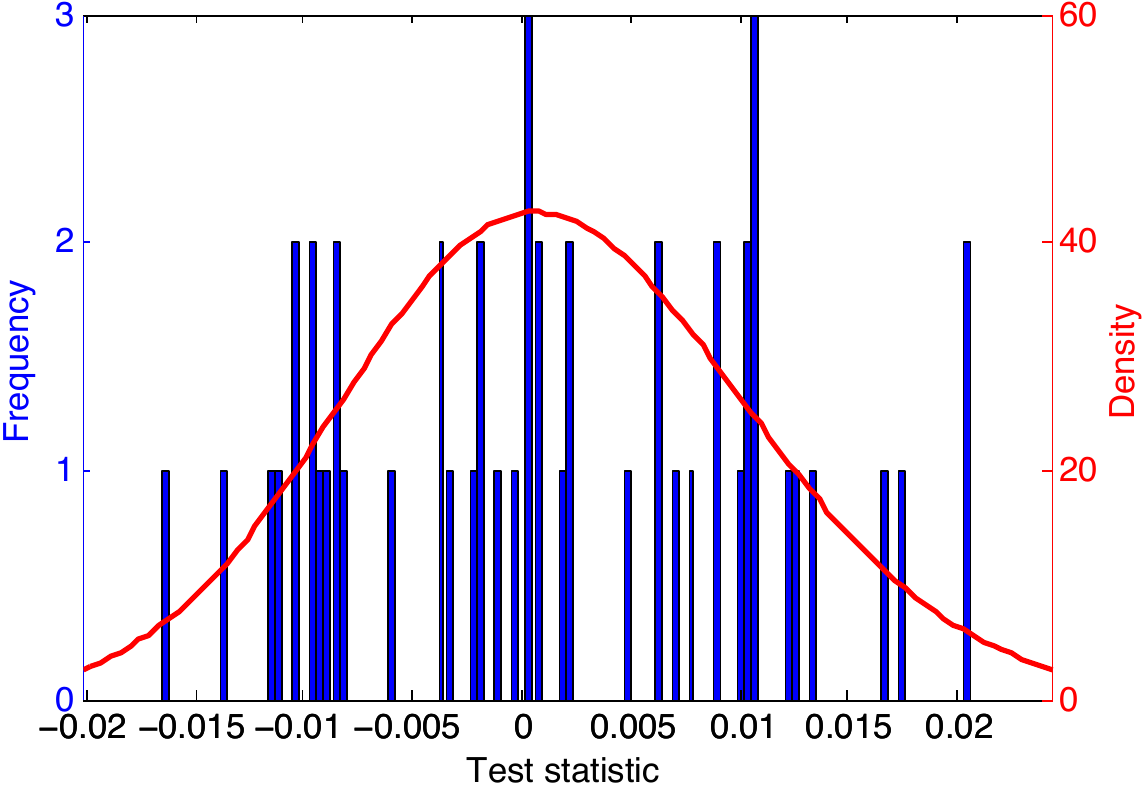}}
	\hspace{10pt}
	\subfloat[$N = 100$]{\includegraphics[height=0.153\textheight]{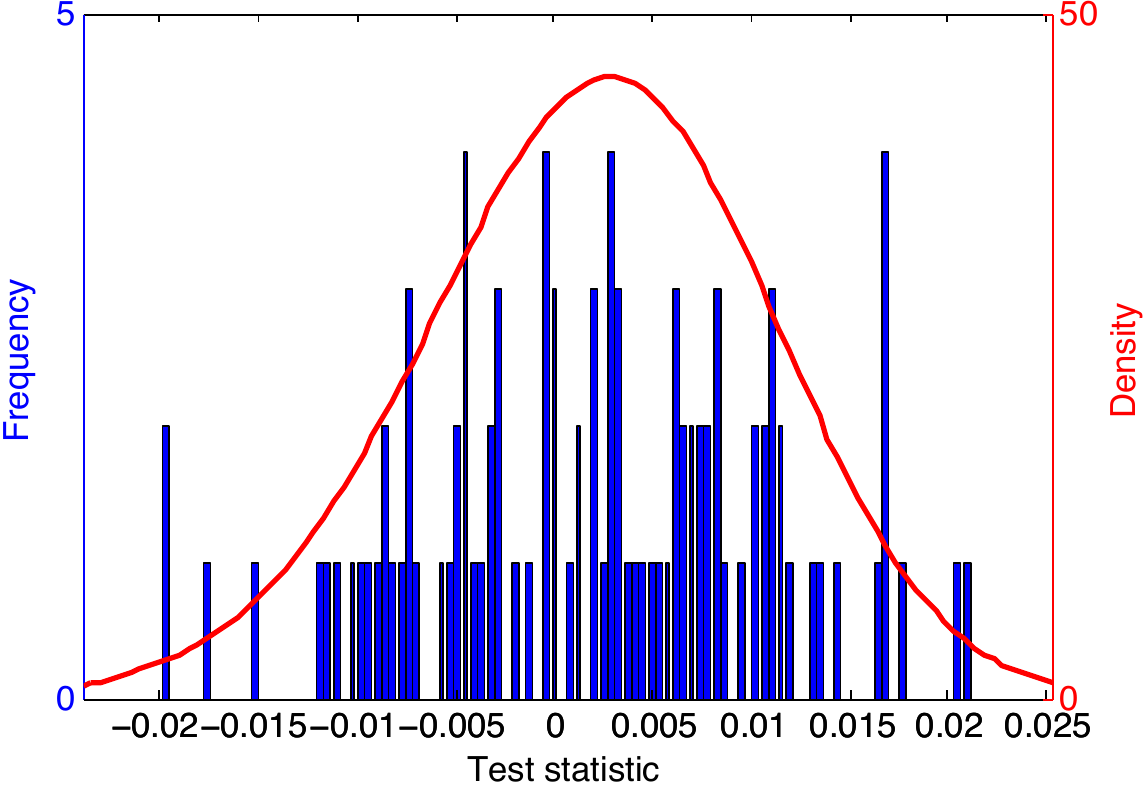}}
	\caption{Example histograms and fitted skew-normal distributions (red curve) after 1000 time steps, for random behaviours with $|A_j|=10$ and $N=10,50,100$. Using score function $z_1$ in test statistic.}
	\label{fig:ra-hist-z1}
\end{figure*}

\begin{figure*}[t]
	\centering
	\subfloat[$N = 10$]{\includegraphics[height=0.153\textheight]{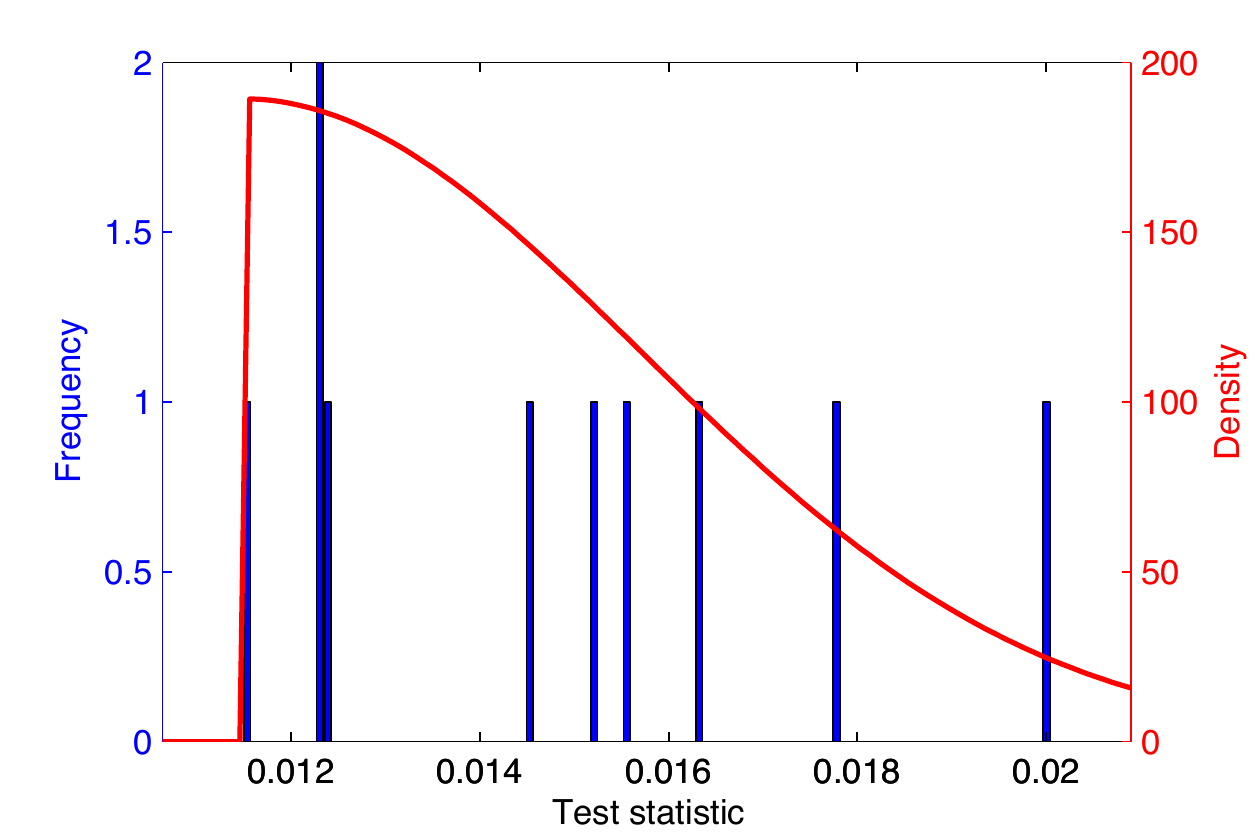}}
	\hspace{10pt}
	\subfloat[$N = 50$]{\includegraphics[height=0.153\textheight]{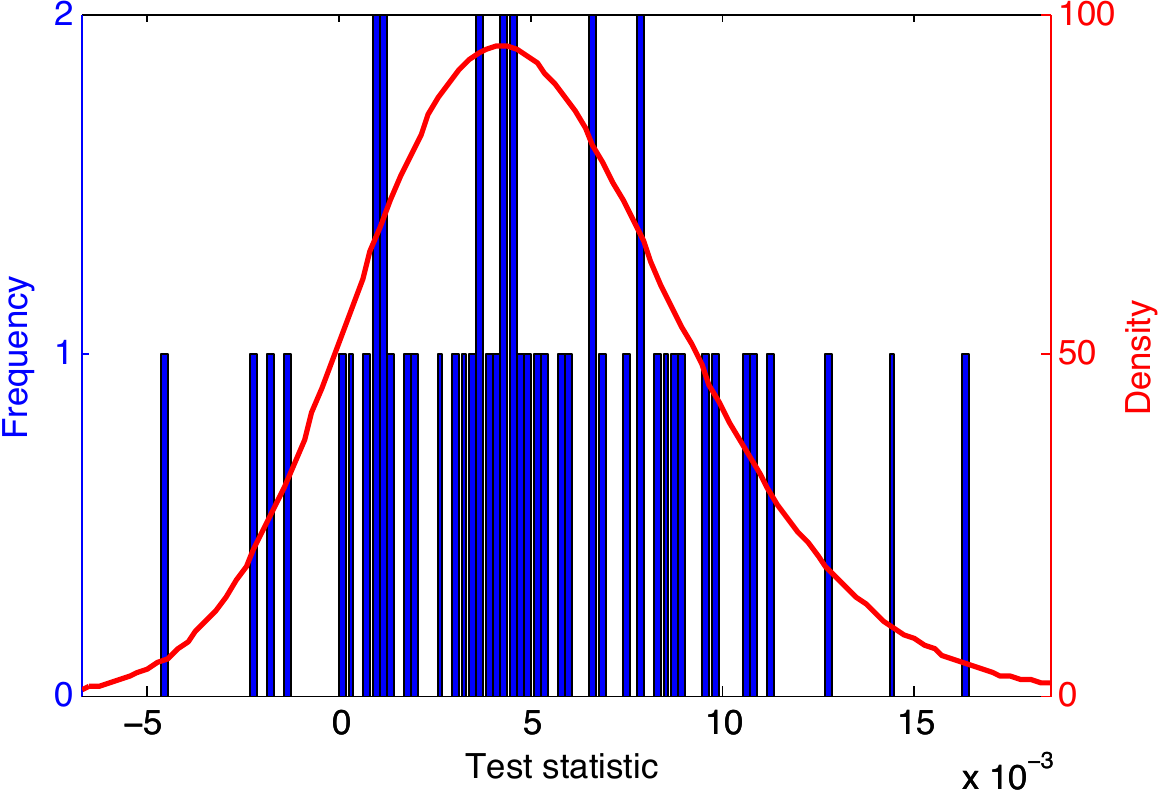}}
	\hspace{10pt}
	\subfloat[$N = 100$]{\includegraphics[height=0.153\textheight]{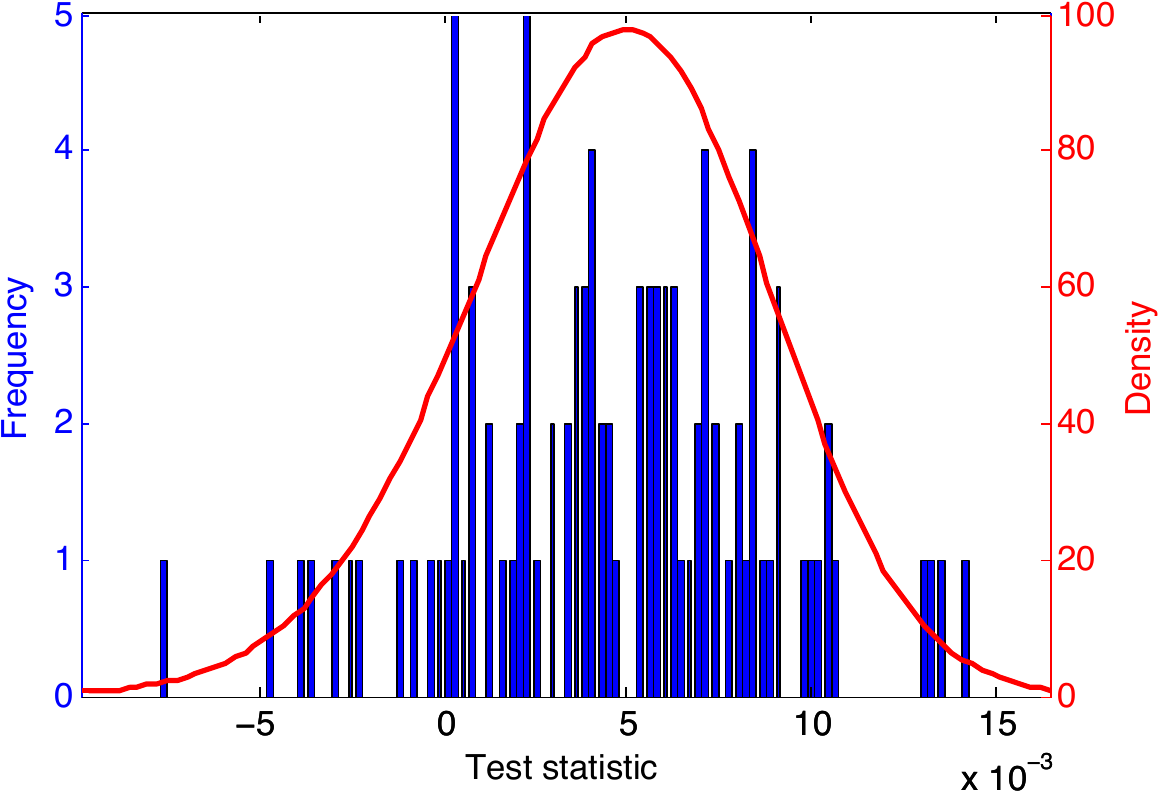}}
	\caption{Example histograms and fitted skew-normal distributions (red curve) after 1000 time steps, for random behaviours with $|A_j|=10$ and $N=10,50,100$. Using score functions $z_1,z_2,z_3$ in test statistic.}
	\label{fig:ra-hist-z123}
\end{figure*}

We also compared the learned skew-normal distributions and found that they fitted the data very well. Figures~\ref{fig:ra-hist-z1} and \ref{fig:ra-hist-z123} show the histograms and fitted skew-normal distributions for two example processes after 1000 time steps. In Figure~\ref{fig:ra-hist-z123}, we deliberately chose an example in which the learned distribution was maximally skewed for $N=10$, which is a sign that $N$ was too small. Nonetheless, in the majority of the processes, the learned distribution was only moderately skewed and our algorithm achieved an average accuracy of 90\% even for $N=10$. Moreover, if one wants to avoid maximally skewed distributions, one can simply restrict the parameter space when fitting the skew-normal (specifically, the shape parameter $\beta$; cf. Section~\ref{sec:learn}).


The flexibility of the skew-normal distribution was particularly useful in the early stages of the interaction, in which the test statistic typically does not follow a normal distribution. Figure~\ref{fig:ra-hist-skew} shows the test distribution for an example process after 10 time steps, using $z_2$ for the test statistic and $N=100$ (the histogram was created using $N=10000$). The learned skew-normal approximated the true test distribution very closely. Note that, in such examples, the normal and Student distributions do not produce good fits.

Our implementation of the algorithm performed all calculations as iterative updates (except for the skew-normal fitting). Hence, it used little (fixed) memory and had very low computation times. For example, using all three score functions and $|A_j|=20, N=100$, one cycle in the algorithm (cf. Algorithm~\ref{alg}) took on average less than 1 millisecond without fitting the skew-normal parameters, and less than 10 milliseconds when fitting the skew-normal parameters (using an off-the-shelf Simplex-optimiser with default parameters). The times were measured using Matlab R2014a on a Unix machine with a 2.6 GHz Intel Core i5 processor.

		\subsection{ADAPTIVE BEHAVIOURS} \label{sec:exp-adapt}

We complemented the ``structure-free'' interaction of random behaviours by conducting analogous experiments with three additional classes of behaviours. Specifically, we used a benchmark framework specified by \citet{acr2015} which consists of 78 distinct $2\hspace{-2pt}\times\hspace{-2pt}2$ matrix games and three methods to automatically generate sets of behaviours for any given game. The three behaviour classes are Leader-Follower-Trigger Agents (LFT), Co-Evolved Decision Trees (CDT), and Co-Evolved Neural Networks (CNN). These classes cover a broad spectrum of possible behaviours, including fully deterministic (CDT), fully stochastic (CNN), and hybrid (LFT) behaviours. Furthermore, all generated behaviours are \emph{adaptive} to varying degrees (i.e. they adapt their action choices based on the other player's choices). We refer to \citet{acr2015} for a more detailed description of these classes (we used the same parameter settings).

The following experiments were performed for each behaviour class, using identical randomisation: For each of the 78 games, we simulated 10 interaction processes, each lasting 10000 time steps. For each process, we randomly sampled behaviours $\pi_i \in \Pi_i, \pi_j \in \Pi_j$ to control agents $i$ and $j$, respectively, where $\Pi_i$, $\Pi_j$ (and $\hysp$) were restricted to the same behaviour class. In half of these processes, we used a correct hypothesis $\hy = \pi_j$, and in the other half, we sampled a random hypothesis $\hy \in \hysp$ with $\hy \neq \pi_j$. As before, we repeated each simulation for $N=10,50,100$ and all constellations of score functions, but found that there were virtually no differences. Hence, in the following, we report results for $N=50$ and the $[z_1,z_2,z_3]$ cluster.

Figure~\ref{fig:lcc-a0} shows the average accuracy achieved by our algorithm for all three behaviour classes. While the accuracy for $\hy = \pi_j$ was generally good, the accuracy for $\hy \neq \pi_j$ was mixed. Note that this was not merely due to the fact that the score functions were imperfect (cf. Section~\ref{sec:test}), since we obtained the same results for all combinations. Rather, this reveals an inherent limitation of our approach, which is that \textit{we do not actively probe aspects of the hypothesis $\hy$}. In other words, our algorithm performs statistical hypothesis tests based only on evidence that was generated by $\pi_i$.

To illustrate this, it is useful to consider the tree structure of behaviours in the CDT class. Each node in a tree $\pi_j$ corresponds to a past action taken by $\pi_i$. Depending on how $\pi_i$ chooses actions, we may only ever see a subset of the entire tree that defines $\pi_j$. However, if our hypothesis $\hy$ differs from $\pi_j$ only in the unseen aspects of $\pi_j$, then there is no way for our algorithm to differentiate the two. Hence the asymmetry in accuracy for $\hy = \pi_j$ and $\hy \neq \pi_j$. Note that this problem did not occur in random behaviours because, there, all aspects are eventually visible.

Following this observation, we repeated the same experiments but restricted $\Pi_i$ to random behaviours, with the goal of exploring $\hy$ more thoroughly. As shown in Figure~\ref{fig:lcc-a1}, this led to significant improvements in accuracy, especially for the CDT class. Nonetheless, choosing actions purely randomly may not be a sufficient probing strategy, hence the accuracy for CNN was still relatively low. For CNN, this was further complicated by the fact that two neural networks $\pi_j,\pi'_j$ may formally be different ($\pi_j \neq \pi'_j$) but have essentially the same action probabilities (with extremely small differences). Hence, in such cases, we would require much more evidence to distinguish the behaviours.

\vspace{-2pt}

	\section{CONCLUSION} \label{sec:conc}

\vspace{-5pt}

We hold the view that if an intelligent agent is to interact effectively with other agents whose behaviours are unknown, it will have to hypothesise what these agents might be doing \emph{and} contemplate the truth of its hypotheses, such that appropriate measures can be taken if they are deemed false. In this spirit, we presented a novel algorithm which decides this question in the form of a frequentist hypothesis test. The algorithm can incorporate multiple statistical criteria into the test statistic and learns the test distribution during the interaction process, with asymptotic correctness guarantees. We presented results from a comprehensive set of experiments, showing that our algorithm achieved high accuracy and scalability at low computational costs.

There are several directions for future work: To bring some structure into the space of score functions, we introduced the concepts of consistency and perfection as minimal and ideal properties. However, more research is needed to understand precisely what properties a useful score function should satisfy, and whether the concept of perfection is feasible or even necessary in the general case. Furthermore, we used uniform weights to combine the computed scores into a test statistic, and we also experimented with alternative weighting schemes to show that the weighting can have a substantial effect on convergence rates. However, further research is required to understand the effect of weights on decision quality and convergence.

Finally, in this work, we assumed that the behaviour of the other agent ($j$) could be described as a function of the information available to our agent ($i$). An important extension would be to also account for information that cannot be deterministically derived from our observations, especially in the context of robotics where observations are often described as random variables.

	\bibliographystyle{plainnat}
	\bibliography{uai15}

\end{document}